\documentclass[12pt]{article}

\usepackage{graphicx,latexsym}
\usepackage{amssymb,amsthm,amsmath,amsbsy}
\usepackage{longtable,booktabs}
\usepackage{mathrsfs}
\usepackage{parskip}
\usepackage[small]{caption}
\usepackage{hyperref}

\newcommand{\dd}{\mathrm{d}}

\newcommand{\scN}{\mathcal{N}}
\newcommand{\ka}{\kappa}
\newcommand{\fts}{\left(\frac{t}{-s}\right)}
\newcommand{\lra}{\relbar\joinrel\longrightarrow}
\newcommand{\reggelimit}{\underset{s \mathrm{\ fixed}}{\underset{t\to\infty}{\lra}}}
\newcommand{\ladder}{\perp\hspace{-2.425 ex}\top}
\newcommand{\fourladder}{\ladder \hspace{-1.5 ex} \ladder\hspace{-1.5 ex} \ladder\hspace{-1.5 ex} \ladder}

\numberwithin{equation}{section}

\begin{document}
\pagestyle{plain}

\begin{flushright} \small{BRX TH-588} \end{flushright}
\vspace*{.1 in}
\begin{center}
{\Large\bf Reggeization of  $\scN$=8 Supergravity \\
and $\scN$=4 Yang--Mills Theory II} \\
\vspace*{.3 in}
{\large Howard J.\ Schnitzer}\footnote{email:
schnitzr@brandeis.edu\\\hspace*{.2in}Supported in part by the DOE
under
grant DE-FG02-92ER40706}\\
Theoretical  Physics Group\\
Martin Fisher School of Physics, Brandeis University\\
Waltham, MA 02454
\end{center}

\vspace*{1.4 in}

\begin{abstract}
The loop expansion for the n-point functions of $\scN=4$ Yang-Mills theory and $\scN=8$ supergravity can be formulated as the loop expansion of scalar field theory with an infinite subclass being the ladder diagrams. We consider the sum of ladder diagrams for gluon-gluon and graviton-graviton scattering in the Regge limit. The reggeization of the gluon and the graviton is discussed in this context and that of \href{http://arxiv.org/abs/hep-th/0701217}{hep-th/0701217}. If the Bern, Dixon, Smirnov conjecture for planar gluon-gluon scattering is correct, then the ladder sum for SU(N) gauge theory at large N, correctly gives the Regge limit, with 
the Regge trajectory function proportional to the cusp anomalous dimension.

In graviton-graviton scattering it is argued that 
the graviton lies on a Regge trajectory.  
Regge cuts are 
also 
present due to infinite sums of non-planar graphs. 
The multiple exchange of Regge poles in non-planar graphs can 
give a countable infinite number of moving Regge cuts which accumulate near $s=0$. 
It is conjectured that this 
may be 
related to the infinite number of non-perturbative massless states which 
remain in the 
limit 
discussed by Green, Ooguri and Schwarz.

\end{abstract}

\newpage

\section{Introduction}

The exciting suggestion\cite{1} that $\scN=8$ supergravity (sugra) might be perturbatively finite in four dimensions ($d=4$) raises several possibilities. \begin{enumerate}
\item If it is not perturbatively finite then $\scN=8$ sugra can only be considered in the context of the compactification of string theory, with the string theory serving as regulator.
\item If $d=4$ , $\scN=8$ sugra is perturbatively finite then there are three possibilities \begin{enumerate}
\item It is a finite theory of quantum gravity, distinct from string theory, or
\item $\scN=8$ sugra in $d=4$ is not a new quantum theory of gravity distinct from string theory. If this is the case, at least in principle one should be able to exhibit all the phenomenon expected of the zero-slope limit of string theory, including the infinite number of zero-mass or low-mass states discussed by 
Green, et.~al.~\cite{2}, 
or 
\item The theory could be perturbatively finite, but non-perturbatively inconsistent. That is, the thoery could lack a non-perturbative completion. As such, the theory could be in 
``swampland'' \cite{2}.
 \end{enumerate}
\end{enumerate}

In previous work \cite{3} it was shown that all of the elementary fields of non-Abelian gauge theories lie on Regge trajectories, with an important example being the gluon 
and quarks 
of QCD. It was also shown that the elementary fields of $\scN=8$ sugra lie on Regge trajectories as well\cite{4,5}. Exploiting the KLT relations\cite{6}, which imply a close relationship between $\scN=8$ sugra and $\scN=4$ YM theory, it was then shown \cite{5} that the factorization conditions for the reggeization of the gluon of $\scN=4$ YM theory implies the reggeization of the graviton of $\scN=8$ sugra if the latter theory is perturbatively finite or, at most logarithmically divergent. Supersymmetry and flavor symmetry then imply that all the elementary fields should reggeize. Thus, if $\scN=8$ sugra is perturbatively finite, it contains Regge trajectories with slope of $\mathcal{O}(\kappa^2)$, consistent with the view that $\scN=8$ sugra could produce the Regge trajectories of perturbative string states.

In \cite{3,4,5} analyticity and Mandelstam counting \cite{7} imply that the reggeization is valid to all orders in perturbation theory, though there is no information on the global properties of the trajectories from perturbative calculations, e.g. the existence of recurrences. The methods of \cite{3,4,5} only give the Regge trajectory and residue to leading order in couplings, though this does not detract from the all-order existence of reggeization mentioned above. It is believed that these leading order results should also be obtainable from the sum of an infinite subclass of diagrams, analogous to the sum of ladder graphs of $\varphi^3$ scalar field theory, keeping only the leading logarithm contributions in the Regge limit. One of the objectives of this paper is to consider the sum of analogous diagrams in leading logarithm approximation for $\scN=8$ sugra and $\scN=4$ YM theory.

It is known that the gluon-gluon scattering amplitude of $\scN=4$ YM theory and graviton-graviton scattering of $\scN=8$ sugra can be expressed schematically as (kinematical factors)$\times$(graphs of scalar field theory)\cite{8}. In both cases one particular infinite subset of graphs is that of $\varphi^3$ ladder graphs. In the limit of large Mandelstam $t$, fixed $s$ (the Regge limit), we show in this paper that in the leading logarithm approximation, one finds $t^{\alpha(s)}$, where $\alpha(s)$ is the Regge trajectory function to leading order, which passes through $J=2$ for the graviton, in perfect agreement with \cite{5}. The discussion of the gluon trajectory, which is more delicate, depending on details of the IR cutoff, is presented in section 2. Also in section 2, we consider the implications of the Bern, Dixon, Smirnov (BDS)\cite{11}, conjecture for the Regge limit of planar gluon-gluon scattering in $\scN=4$ YM theory.

In theories with Regge poles, there are also moving Regge cuts. Such Regge cuts cannot appear from sums of planar graphs \cite{9,10}, so that any Regge cuts are suppressed by at least $\mathcal{O}(1/N)$ in large N, SU(N), $\scN=4$ YM theory. There is no such suppression, however, in $\scN=8$ sugra, so there \emph{must} be Regge cuts on the same footing as the Regge poles of $\scN=8$ sugra. What is their interpretation in terms of the zero-slope limit of string theory, if any? In section 3 we discuss reggeization of $\scN=8$ sugra in the context of ladder sums for graviton-graviton scattering. The possible structure of the moving Regge cuts of the theory is contained in section 4. We then make the \emph{speculation} that the moving Regge cuts of $\scN=8$ sugra are related to the infinite number of zero-mass states discussed by Green et. al.\cite{2} In section 5, we summarize our findings and offer conclusions.

\section{Regge Behavior of $\scN=4$ YM Theory}

\begin{figure}[!h!]
  \includegraphics[width=\textwidth]{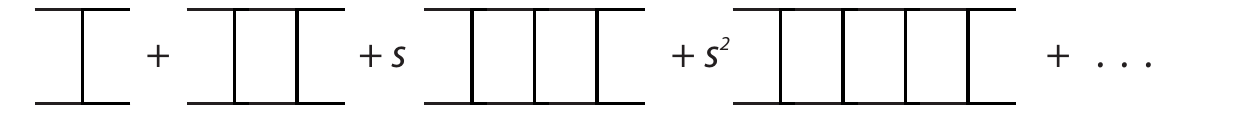}\\
  \caption{}\label{lad1}
\end{figure}

The loop expansion for the 4-point functions of $\scN=4$ Yang-Mills theory can be formulated as a loop expansion of scalar field theory. An infinite subclass of terms in the expansion are ladder graphs. The $s$-channel ladders \cite{8} for $\scN=4$ YM theory are shown in fig. \ref{lad1}.

The leading color amplitude corresponding to fig. \ref{lad1} is \cite{8}
\begin{equation}
A_4^{\mathrm{ladder}} = [s t A_{tree}] \sum_{n=1}^\infty [ s^{n-2} \underbrace{\fourladder}_{n-\mathrm{rungs}} ]
\label{lad2}
\end{equation}
with conventions as in \cite{4,5} and helicities as shown;
\begin{equation}
A_4 (1,2;3,4)=A_{tree}(1,-1;1,-1)=4\left(\frac{-t}{s}\right)^2\ s\ [ \frac{\alpha}{u} + \frac{\beta}{t} ]g^2
\end{equation}
where
\begin{equation}
\alpha-\beta = f_{abg} f_{c d g} = [T_c,T_d]_{ab}
\label{fag}
\end{equation}
and
\begin{equation}
\alpha+\beta = \{T_c,T_d\}_{ab}
\end{equation}
and
\begin{eqnarray} \nonumber
s&=&(k_1+k_2)^2 =-2q^2 (1-\cos \theta)\\ \nonumber
t&=&(k_1+k_4)^2 = 4q^2 \\
u&=&(k_1+k_3)^2 = -2 q^2 (1+\cos \theta)
\end{eqnarray}
are Mandelstam variables, $q$ is the CM momentum and $\theta$ is the scattering angle.

In the Regge limit, $t\to\infty$, $s$ fixed
\begin{equation}
A_{tree} \reggelimit \ \ 4 \fts(\alpha-\beta) g^2
\end{equation}
which is in the adjoint representation. Similarly, the one-loop amplitude is
\begin{equation}
(A_4)_{1-\mathrm{loop}} = \lambda\ t^2\ I_4^{1-\mathrm{loop}}(s,t)
\end{equation}
where $\lambda = g^2 N$ is 't Hooft coupling and \cite{8,10}
\begin{eqnarray} \label{i4one}
I_4^{1-\mathrm{loop}} &=& \int \frac{\dd^4 p}{(2 \pi)^4} \frac{1}{\left[p^2 (p-k_1)^2(p-k_1-k_2)^2(p+k_4)^2\right]} \\
&\reggelimit& \frac{1}{16 \pi^2}\frac{K(s)}{t}\log \fts
\end{eqnarray}
where \cite{10}
\begin{equation}
K(s)=-\int_0^1 \dd \beta_1 \dd \beta_2 \frac{\delta(\beta_1+\beta_2-1)}{\left[\beta_1 \beta_2 s - (\beta_1+\beta_2)\mu^2\right]}
\label{IRc}
\end{equation}
with $\mu^2$ an IR cutoff. For $(-s) >>\mu^2$
\begin{equation}
K(s) \underset{(-s)>>\mu^2}{\lra} 
-\frac{2}{s} \log \left( -s \over \mu^2 \right) 
+ \mathcal{O}(\mu^2/s).
\label{K} 
\end{equation}
Then the ladder sum (\ref{lad2}) becomes, in the Regge limit,
\begin{eqnarray} \nonumber
A_4^{ladder} &=& -g^2 (t^2) \left(\frac{1}{t}\right) \sum_{n=1}^\infty s^{n-2} \frac{\left[ \left(\frac{\lambda}{16 \pi^2}\right) K(s) \log \fts\right]^{n-1}}{(n-1)!} \\ \nonumber
&=& g^2 \fts \exp \left[ \frac{\lambda}{16 \pi^2} s\ K(s) \log \fts \right] \\ \label{laddamp}
&=& g^2 \fts^{\alpha(s)}
\label{ladder} 
\end{eqnarray}
where
\begin{equation}
\alpha(s) = 1 + \frac{\lambda}{16 \pi^2} s K(s) + \mathcal{O}(\lambda^2)
\label{alp}
\end{equation}

Since (\ref{fag}) is in the adjoint representation in the s-channel, (\ref{alp}) is the gluon Regge trajectory to leading order in $\lambda$. 
Given the IR cutoff as in (\ref{IRc})
\begin{equation}
\alpha(s) \underset{(-s)>>\mu^2}{\lra} 
1 - \frac{\lambda}{8\pi^2} \log \left( -s \over \mu^2 \right) 
+ \mathcal{O}(\lambda^2)  
\end{equation}
while
\begin{equation}
K(0)=\frac{1}{\mu^2}
\end{equation}
so that
\begin{equation}
\alpha(s) \underset{s\to0}{\lra} 1 + \frac{s}{16 \pi^2 \mu^2} + \cdots
\label{tttt}
\end{equation}
Thus, the Regge trajectory function passes through $j=1$ at $s=0$. That is, the gluon reggeizes, with the IR regulator as in (\ref{IRc}). The Regge trajectory, however, depends sensitively on the IR cutoff. We have chosen the regulator to agree with the reggeization in \cite{4,5}.

There is a subtlety in the Mandelstam counting argument \cite{7}. In addition to renormalizability, one also requires the absence of IR singularities, or the presence of a mass gap, which the regulator (\ref{IRc}) provides. Thus, with this regulator, the gluon Regge trajectory satisfies $\alpha(0)=1$, as seen in (\ref{tttt}), consistent with the analysis of \cite{4,5}.

What about the contribution of other planar graphs computed in the Regge limit, which are not included in (\ref{laddamp})? The most interesting possibility is provided by the BDS conjecture \cite{11} 
Then 
we suggest that (\ref{ladder})  
is replaced by
\begin{eqnarray} \nonumber
A_4 &=& g^2 \fts \exp\left[\frac{f(\lambda)}{8} s K(s) \log \fts + C(\lambda) \right] \\
&=& g^2 \beta(\lambda) \fts^{\alpha(\lambda)}
\end{eqnarray}
where $f(\lambda)$ is the cusp anomalous dimension. 
The trajectory function becomes, for $(-s)>>\mu^2$
\begin{equation}
\alpha(\lambda) = 
1 - \frac{f(\lambda)}{4} \log \left( -s \over \mu^2 \right) 
\label{al}
\end{equation}
to all orders in perturbation theory. It is known that\cite{12,13}
\begin{equation}
f(\lambda) = \left\{ \begin{array}{cc}
8\left(\frac{\lambda}{16 \pi^2}\right) - 16\ \zeta_2 \left(\frac{\lambda}{16 \pi^2}\right)^2 + \cdots  &    \mbox{ in weak coupling }\\ \\
\frac{\sqrt{\lambda}}{\pi} - \frac{3 \log 2}{4\pi} + \mathcal{O}\left(\frac{1}{\sqrt{\lambda}}\right)  &    \mbox{ in strong coupling }
\end{array} \right.
\end{equation}
However, to repeat, whether $\alpha(s=0)=1$ or not, is an IR issue.

Another issue is that (\ref{laddamp}) is only the \emph{leading} term in the Mellin transform of the box diagram. There are subleading terms in powers of $(1/t)$ related to (\ref{i4one}) which are \emph{daughter} trajectories of the gluon trajectory. Thus, (\ref{laddamp}) \emph{also} has a sequence of daughters as well. The issue of the first daughter becomes important when we consider Regge trajectories for graviton-graviton scattering in $\scN=8$ sugra.

Finally, note that moving Regge cuts involve non-planar graphs \cite{9,10}, so any possibly moving Regge cuts in SU(N) $\scN=4$ YM are suppressed by at least $\mathcal{O}(1/N)$. There is no such suppression in $\scN=8$ sugra, as discussed in section 4.

\section{Reggeization of $\scN=8$ sugra}

The KLT relations 
\cite{6} 
assist in recasting the loop expansion for graviton-graviton scattering in terms of scalar field theory. The non-bubble or triangle hypothesis, which has been checked up to four loops \cite{1}, suggests that $\scN=8$ sugra in $d=4$ may be no more divergent than $\scN=4$ YM, i.e., perturbatively finite. Since the theory is not conformal, IR is not an issue for our calculations as we shall see explicitly in the sum of ladder graphs.

The structure of the ladder graphs in $\scN=8$ sugra is similar, but not identical, to that of $\scN=4$ YM. The KLT relations 
allow us to write the tree amplitude for graviton-graviton scattering in terms of gluon-gluon scattering. That is\cite{8}
\begin{equation}
(s t u) M_4^{tree}(1,2;3,4) = \left( \frac{\ka}{2} \right)^2 \left( s t \left[ A_4^{tree}(1,2;3,4)\right]^2 \right)
\end{equation}
which, from (\ref{lad2}), implies that
\begin{equation}
M_4^{tree}(1,2;3,4) = \frac{\left(\frac{\ka}{2}\right)^2 t^4}{s t u}
\end{equation}

Similarly, the s-channel ladders have a structure similar to that of fig. \ref{lad1} and (\ref{laddamp}), but instead with factor $s^{2L-2}$ for L-loops \cite{8}. That is, for $\scN=8$ sugra the s-channel ladder contribution in the Regge limit is, for $L\geq1$ loops, up to an overall constant
\begin{eqnarray} \nonumber
M_4^{L\geq 1}(1,2;3,4) &\rightarrow& \sum_{L=1}^\infty \left[ \left(\frac{\ka}{2}\right)^2 s \right]^{L+1} \fts^3 \frac{\left[ s K(s) \log\fts \right]^L}{L!} \\
&=& \left[ \left(\frac{\ka}{2}\right)^2 s \right] \fts^3 \left[ \fts^{\bar\alpha(s)} -1 \right]
\end{eqnarray}
where
\begin{equation}
\bar\alpha(s) = \left[\left(\frac{\ka}{2}\right)^2 s \right] s K(s)
\label{t}
\end{equation}

and $K(s)$ is identical to (\ref{IRc}). Adding this to the tree amplitude
\begin{equation}
M_4^{ladder}(s,t) \reggelimit \left[ \left(\frac{\ka}{2}\right)^2 s \right] \fts^2 + \left[ \left(\frac{\ka}{2}\right)^2 s \right] \fts^3 \left[ \fts^{\bar\alpha(s)} -1 \right]
\label{m4lad}
\end{equation}

Before drawing any conclusions about the Regge behavior, we must deal with two subtleties which were not relevant for our discussion of gluon-gluon scattering, as they did no affect the conclusion regarding the gluon reggeization. Here we must consider both the \emph{signatured amplitudes} and \emph{daughter trajectories}. After considering ladders for the right signature amplitudes, we will find that the graviton lies in a trajectory passing through $J=2$. There is also a wrong signature amplitude passing through $J=3$, but with zero Regge residue, as can already be seen in (\ref{m4lad}).

The signatured amplitudes \cite{4} for the ladders, which contribute in the Regge limit are
\begin{equation}
\left[ M_4^{ladder}(s,t) \pm M_4^{ladder}(s,u) \right] \equiv M_4^\pm(s,t)
\end{equation}
which \emph{includes} daughters, as we will discuss, with the $+ (-)$ sign for the right (wrong) signature amplitude.

If one computes the Regge limit of (\ref{i4one}) using the Mellin transform \cite{9,10}, then one finds a sequence of daughters displaced by negative integers from the leading trajectory, but with the same Regge residue. 
Exhibiting only the leading trajectory and first daughter, 
because the daughter and parent have opposite signature,
the right signature amplitude becomes
\begin{eqnarray}
\label{monster}
M_4^+(s,t) &\to& \left[ \left(\frac{\ka}{2}\right)^2 s \right] \left[ \fts^2+\left(\frac{u}{s}\right)^2\right] \\ \nonumber
&+& \left[ \left(\frac{\ka}{2}\right)^2 s \right] \Biggl\{ \fts^3 \left[ \fts^{\bar\alpha(s)}-1\right]
-  \left(\frac{u}{s}\right)^3 \left[\left(\frac{u}{s}\right)^{\bar\alpha(s)}-1 \right] \Biggr\} \\ \nonumber
&+&  \left[ \left(\frac{\ka}{2}\right)^2 s \right] \Biggl\{ \fts^2 \left[ \fts^{\bar\alpha(s)} - 1 \right]
+ \left(\frac{u}{s}\right)^2 \left[ \left(\frac{u}{s}\right)^{\bar\alpha(s)} -1 \right] \Biggr\}
\end{eqnarray}

Since $s+t+u=0$, (\ref{monster}) becomes, in the Regge limit,
\begin{equation} \label{monst2}
M_4^+(s,t) \reggelimit  2 \left[\left(\frac{\ka}{2}\right)^2 s \right] \fts^{\alpha(s)}  +  \cdots
\end{equation}
where
\begin{eqnarray} \nonumber
\alpha(s) &=& 2 + \bar\alpha(s) \\
\alpha(0) &=& 2 \label{rt2}
\end{eqnarray}
since
\begin{equation}
\left[ s^2 K(s) \right]_{s=0} = 0
\end{equation}
Therefore the IR divergence of the trajectories is not an issue for this calculation. The interpretation of (\ref{monst2}) is that the graviton has reggeized, turning into a moving Regge pole passing through $J=2$, with Regge trajectory given by (\ref{rt2}), (\ref{t}) and (\ref{IRc}). This trajectory is approximately linear with slope $\mathcal{O}(\kappa^2)$. The wrong signature amplitude, $M_4^-(s,t)$, is that of a trajectory passing through $J=3$, but with vanishing residue so that there \emph{no massless particles} with $J=3$. Results for both signatured amplitudes are in agreement with \cite{4,5}.

It would be interesting if there was an analogue of the BDS \cite{11} conjecture for the planar diagram contributions to graviton-graviton scattering, which would then address possible corrections to (\ref{monst2}).

\section{Regge cuts in $\scN=8$ supergravity}

This is a speculative discussion concerning the presence of moving Regge cuts in $\scN=8$ sugra and their possible relationship to string theory. Moving cuts on the physical sheet of the scattering amplitude can only come from non-planar diagrams. An example of the contribution of the exchange of two s-channel ladders, in scalar field theory, forming a non-planar diagram gives\cite{9,10}
\begin{equation}
\int_{\lambda\leq 0} \dd s_1 \dd s_2 \frac{\left[f(s,s_2,s_2)\right]^2}{\lambda^{1/2}(s,s_1,s_2)} t^{\alpha(s_1)+\alpha(s_2) -1}
\end{equation}
where $f(s,s_1,s_2)$ is essentially the Regge residue and
\begin{equation}
\lambda(s,s_1,s_2) = s^2 +s_1^2+s_2^2 - 2 s s_1 - 2 s s_2 -2 s_1 s_2
\end{equation}
In terms of the complex angular momentum plane $j$, this is
\begin{equation}
\int \dd s_1 \dd s_2 \frac{f^2}{\lambda^{1/2}} \frac{1}{\left[j-\alpha(s_1)-\alpha(s_2) +1\right]}
\label{42}
\end{equation}
The poles at $\alpha(s)=0$ give a singularity in (\ref{42}) at $j=1$, where the position of the cut depends on $\alpha(s_1)+\alpha(s_2)-1$ in the region of integration. If one assumes that $\alpha(s)$ depends linearly on $s$, as is the case in $\scN=8$ sugra, then the singularity occurs at $s_1=s_2$ on the boundary $\lambda=0$, i.e. $s_1=s_2=s/4$. Thus \cite{10}
\begin{eqnarray}\nonumber
j&=& 2\ \alpha\left(\frac{s}{4}\right) - 1 \\
&=& 2 \left[ \alpha(0) + s \alpha'(s)\right] - 1
\end{eqnarray}
As an example, if $\alpha(0)=1$, then the cut also passes through $j=1$ when $s=0$.

Instead of exchanging two Regge poles to obtain a cut, one may also exchange a cut and a pole, or a pair of cuts. This procedure generates a sequence of singularities at positions\cite{10}
\begin{equation}
j = n\ \alpha\left(\frac{s}{n^2}\right) - (n-1) \qquad \qquad n=1,2,3 \cdots
\label{44}
\end{equation}
which all pass through $j=1$ when $s=0$. Since
\begin{equation}
\alpha(s) = 1 + \alpha'(0)\ s
\end{equation} in this example, the singularities (\ref{44}) only extend a finite distance to the right for any fixed $s$, in units of $\ka^2$ for $\scN=8$ sugra.

The existence of Regge poles in $\scN=8$ sugra thus suggests the existence of an infinite number of Regge cuts in theory, each with singularities $0\leq s\leq \mathcal{O}(1/\ka^2)$. A possibility worth exploring is that these cuts are associated to the zero-mass or low-mass states discussed by Green, et. al.\cite{2}, in obtaining $\scN=8$ sugra in the decoupling limit from string theory, although this may be a very difficult task to verify in detail. 

\section{Concluding Remarks}

The contribution of the Regge limit of the sum of ladder graphs for both gluon-gluon scatter in $\scN=4$ YM and graviton-graviton scattering in $\scN=8$ sugra was the principle focus of this paper. It was shown that the graviton lies on a moving Regge trajectory passing through $J=2$, in agreement with the analyticity analysis of \cite{5}. However, the issue of the reggeization of the gluon is more delicate, as discussed in section 2. 
Accepting the BDS conjecture\cite{11} for planar gluon-gluon scattering, 
we 
argued 
that the ladder sum gives the 
Regge limit for 
the amplitude, but with the replacement of the coupling in the Regge trajectory function by the cusp anomalous dimension.

The implications of the reggeization of the graviton and other fundamental fields of $\scN=8$ sugra is that the perturbative states of string theory can plausibly emerge within the context of the sum of graphs of $\scN=8$ sugra. Among these graphs are also infinite sums of non-planar graphs, which will lead to Regge cuts. Multiple exchange of Regge poles or cuts can give an infinite number of moving Regge cuts in scattering processes, both charged and uncharged, which accumulate near $s=0$. We conjecture that these are related to the infinite number of massless or low-mass states discussed by Green et. al.\cite{2}

If $\scN=8$ sugra is indeed perturbatively finite, the central issue pivots on possibilities (2b) or (2c) of the introduction. That is, 
can all the phenomena of the 
compactified string theory be plausibly exhibited (at least qualitatively) starting with $\scN=8$ sugra, and is such a theory non-perturbatively consistent? These are important questions which deserve further consideration.

\section*{Acknowledgements}
We wish to thank 
Jacques Distler, Mike Duff,    
Albion Lawrence, Steve Naculich, Warren Siegel, 
Kelly Stelle, Paul Townsend,    
and Edward Witten for conversations and 
Oren Elrad for help in preparing the manuscript.
We also thank Steve Naculich for correcting the error in (\ref{K}) in 
version 1 of this paper.


\end{document}